# EFFECT OF He$^+$ IRRADIATION ON Fe-Cr ALLOYS: MÖSSBAUER–EFFECT STUDY[*]

S. M. Dubiel[1*], J. Cieślak[1], H. Reuther[2]

[1]AGH University of Science and Technology, Faculty of Physics and Applied Computer Science, PL-30-059 Kraków, Poland, [2]Helmholtz-Zentrum Dresden-Rossendorf, Postfach 510119, D-01314 Dresden, Germany

**Abstract**

Effect of He ions irradiation on three model $Fe_{100-x}Cr_x$ alloys ($x$ =5.8, 10.75 and 15.15) was investigated with the conversion electron Mössbauer spectroscopy. The study of the alloys irradiated with 25 keV ions revealed that the strongest effect occured in the $Fe_{84.85}Cr_{15.15}$ sample where an inversion of a short-range-order (SRO) parameter was found. Consequently, the investigation of the influence of the irradiation dose, $D$, was carried out on the chromium-most concentrated sample showing that the average hyperfine field, $<B>$, the average angle between the normal to the sample's surface and the magnetization vector, $<\Theta>$, as well as the actual distribution of Fe/Cr atoms, as expressed by SRO parameters, strongly depend on $D$. In particular: (a) $<B>$ increases with $D$, and its maximum increase corresponds to a decrease of Cr content within the two-shell volume around the probe $^{57}$Fe nuclei by ~2.3 at%, $<\Theta>$ decreases by ~13 degree at maximum, (c) SRO-parameter averaged over the two-shell volume increases with $D$ from weakly negative value (indicative of Cr atom clustering) to weakly positive value (indicative of Cr atoms ordering). The inversion takes place at $D \approx 7$ dpa.

* Corresponding author: Stanislaw.Dubiel@fis.agh.edu.pl

# 1. Introduction

Chromium based ferritic steels (FS) constitute an important class of structural materials. Their technological importance follows from their very good swelling, high temperature corrosion and creep resistance properties [1-2]. Consequently, FS are regarded as appropriate materials to be applied for the new generation of nuclear power facilities such as generation IV fission reactors and fusion reactors as well as for other technologically important plants like high power spallation targets [3-5]. In particular, they are used for a construction of such systems as fuel cladding, container of the spallation target or primary vessel . These devices work at service not only at elevated temperatures but also under irradiation conditions. In these conditions, the materials undergo irradiation damage that can seriously deteriorate their mechanical properties. On the lattice scale, the radiation causes lattice defects, and, consequently, a redistribution of Fe/Cr atoms that can result in a short- range order (SRO) or phase decomposition into Fe-rich and Cr-rich phases. Both these effects result, among other, in an enhancement of embrittlement. A better knowledge of the effect of irradiation on the useful properties of FS and underlying mechanisms is an important issue as it may help to significantly improve their properties. Fe-Cr alloys, that constitute the basic ingredient of FS, are often used as model alloys for investigations of both physical and technological properies [6 and references therein] .

In this paper the effect of $He^+$ irradiation on three model (EFDA/EURATOM) $Fe_{100-x}Cr_x$ of a practical importance, as a production of helium occurs during exposure of the devices to proton and/or neutron irradiation [1]. Its presence has a negative effect on the mechanical properties. In particular, it lowers the critical stress for intergranular structure and may induce a severe decrease of the fracture toughness [7]. Therefore, the understanding not only of the radiation damage but also the effect of helium on the mechanical properties of FS is one of important topics to be addressed.

## 2. Experimental

### 2. 1. Samples

Samples investigated in this study were prepared from EFDA/EURATOM model Fe-Cr alloys that had been fabricated in 2007. They were delivered in the form of bars 10.9 mm in diameter, in a re-crystallized state after cold reduction of 70% and then heat treated for 1h under pure argon flow at the following temperatures: 750 °C for $Fe_{94.2}Cr_{5.8}$, 800 °C for $Fe_{89.25}Cr_{10.75}$ and 850 °C for $Fe_{84.85}Cr_{15.15}$ followed by air cooling. For the Mössbauer spectroscopic (MS) measurements, a slice ~1 mm thick was cut off from each bar using a diamond saw, and it was subsequently cold-rolled (CR) down to a final thickness of 20-30 μm. A list of the investigated samples is displayed in Table 1.

The samples were irradiated with 25 keV He-ions with the Facilities at the Ion Beam Center of the Helmholtz-Zentrum Dresden-Rossendorf, Germany. Final doses of the irradiation are given in Table 1.

### 2. 2. Measurements

Because of the implantation depth that in the present case was ≤ ~0.225 μm, the Mössbauer spectra were measured at room temperature (RT) recording conversion electrons in backscattering geometry with a conventional constant acceleration spectrometer and a

57Co(Rh) source of 14.4 keV gamma-rays with an activity of nominally 3.7 GBq. Samples were built in a proportional gas flow counter with a He/methan mixture as counting gas. Examples of the spectra measured for the samples No 1- 4 are presented in Fig. 1

**Table 1**
Investigated model (EFDA/EURATOM) samples of $Fe_{100-x}Cr_x$ alloys. CR stays for the cold-rolled state.

| No | x | Metallurgical state | Dose [$He^+/cm^2$] | Dose [dpa] |
|---|---|---|---|---|
| 1 | 5.8 | CR | 0 | 0 |
| 2 | 5.8 | CR | $1.2 \cdot 10^{17}$ | 7.5 |
| 3 | 10.75 | CR | 0 | 0 |
| 4 | 10.75 | CR | $1.2 \cdot 10^{17}$ | 7.5 |
| 5 | 15.15 | CR | 0 | 0 |
| 6 | 15.15 | CR | $6 \cdot 10^{16}$ | 4 |
| 7 | 15.15 | CR | $1.2 \cdot 10^{17}$ | 7.5 |
| 8 | 15.15 | CR | $4 \cdot 10^{17}$ | 25 |

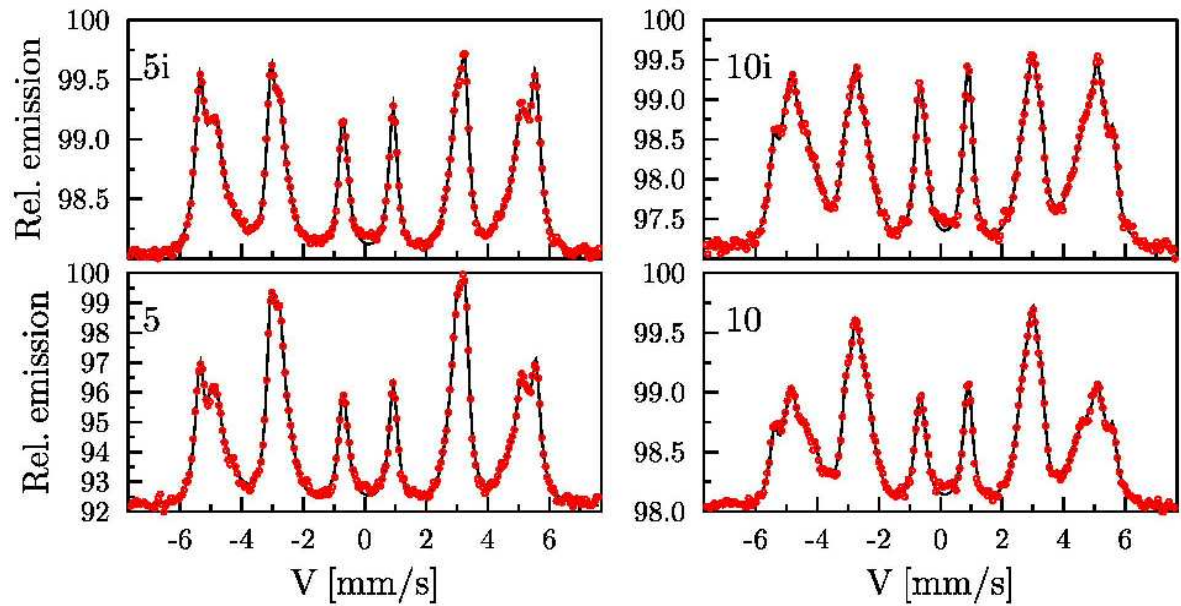

**Fig. 1** $^{57}$Fe CEMS spectra recorded at RT on (a) non-irradiated, and (b) irradiated $Fe_{100-x}Cr_x$ samples (left x =5.8) and (right x =10.75). Index "i" marks the irradiated samples.
3

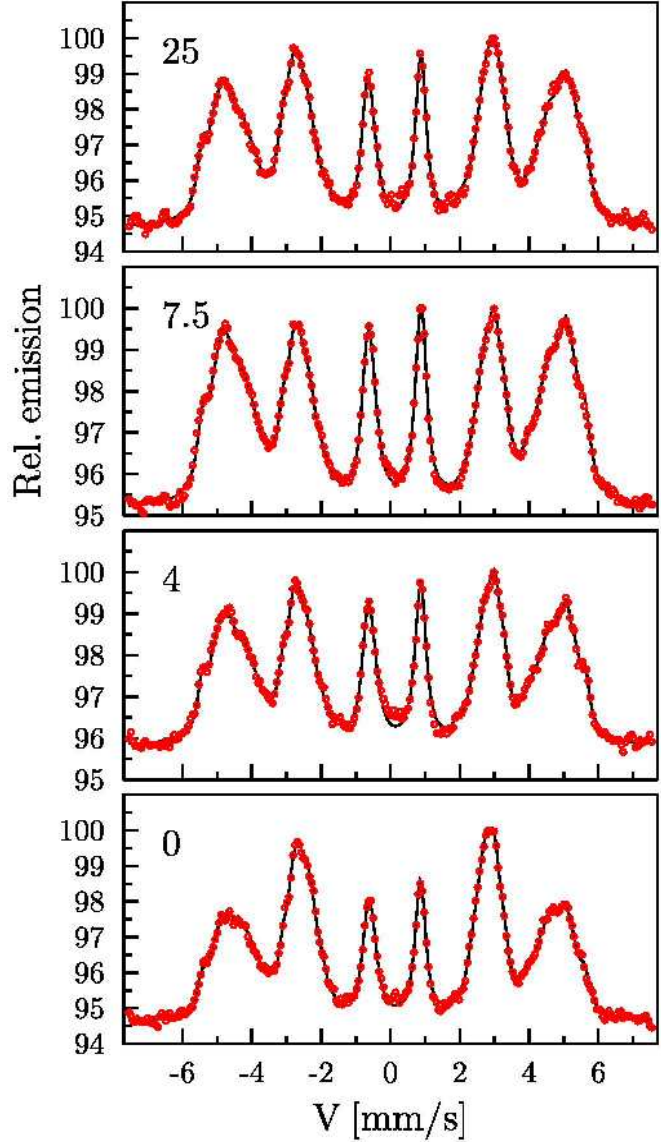

**Fig. 2** $^{57}$Fe CEMS spectra recorded at RT on Fe$_{84.85}$Cr$_{15.15}$. The irradiation dose of He$^+$ (in dpa) is shown as a label for each spectrum.

## 2. 3. Spectra analysis

As one of the aims of this study it was to investigate the effect of irradiation on Fe/Cr atoms distribution, the spectra were analyzed in terms of a superposition method and in the two neighbour-shells approximation as described in detail elsewhere [4]. In particular, we assumed that Cr atoms situated within the first (*1NN*) and the second (*2NN*) neighbour-shell around the probe $^{57}$Fe nuclei had a measurable effect on the spectral hyperfine parameters viz. the hyperfine field (*H*) and the isomer shift (*IS*). Using the *1NN-2NN* model, we further assumed that the effect of Cr atoms on *H* and *IS* was additive i.e. $X(m,n) = X(0,0) + m·\Delta X1 + n·\Delta X2$, where *X=H or IS*, *ΔXk* is a change of the hyperfine field (*X=H*) or that of the isomer shift (*X=IS*) due to one Cr atom situated in *1NN* (*k=1*) or in *2NN* (*k=2*). The number of Cr atoms is indicated by *m* for *1NN* and by *n* for *2NN*. All the spectra recorded on the Fe$_{84.85}$Cr$_{15.15}$ sample were fitted simultaneously with the following common parameters: *X(0,0)* and *ΔXk*. The following parameters were free for each spectrum: *G1, G2, G3 (line* *widths)*, *C2* (Clebsch-Gordon coefficient of the second/fifth line in a sextet), *P(m,n)* (probability of finding *m* Cr atoms in *1NN* and *n* ones in *2NN*). For a binary alloy with a random distribution of atoms *P(m,n)* – values can be found using the binomial distribution. Within the *1NN-2NN* approximation, there are *N* = 63 possible configurations, *(m,n)*. Foretunately, Most of them have vanishingly small probabilities, and can therefore be neglected. In practice, one usually takes into account only the most probable ones in order to fulfill the condition: *P(m, n; x)* ≥ 0.99. This condition significantly reduces *N* from 63 to e. g. *N* = 4 (*x* = 1), *N* = 10 (*x* = 10), *N* = 14 (*x* = 15).

Based on *P(m,n)*-values obtained from the analysis of the spectra, the average numbers of Cr atoms in *1NN*, *<m>* and that in *2NN*, *<n>*, were calculated, and next used to determine short-range order (SRO) parameters *<$\alpha_1$>*, *<$\alpha_2$>* and *<$\alpha_{12}$>* for *1NN*, *2NN*, and *1NN-2NN*, respectively. The effect of the irradiation dose on the SRO-parameters and on the spectral parameters is presented and discussed below.

## 3. Results and discussion

### 3. 1. Spectral parameters

The above-described fitting procedure has been used to successfully analyze all the measured spectra. In the case of the $Fe_{84.85}Cr_{15.15}$ sample, the common parameters had the following values: *B(0,0)=34.31 T, $\Delta B_1$=-30.96 T, $\Delta B_2$=-19.06 T, I(0,0)=0.093mm/s, $\Delta IS_1$=-0.020 mm/s and $\Delta IS_1$=-0.010 mm/s*. These values are very close to those obtained previously for the Fe-Cr alloy of similar composition [8,9]. The most sensitive to the irradiation spectral parameter turned out to be the average hyperfine field, *<B>*, that was calculated as a weighted over the hyperfine fields associated with particular atomic configurations, $<B> = \sum_{m,n} P(m,n)B(m,n)$, where *B(m,n)* is the hyperfine field for *(m,n)* atomic configuration.

### 3. 2. Effect of irradiation

#### 3. 2. 1. Samples with different composition

In order to study whether or not the effect of irradiation depends on the sample's composition, the $Fe_{100-x}Cr_x$ samples with x = 5.8, 10.75 and 15.15 were irradiated with He ions to the final dose of $1.2 \cdot 10^{17}$ $He^+/cm^2$ (7.5 dpa). CEMS spectra recorded at RT before the irradiation and afterwards are displayed in Fig. 1a. They were analyzed in terms of the fitting procedure described above. The most-efected spectral parameter turned out to be *C2*, hence the angle between the direction of the incident γ-rays (perpendicular to the sample's surface) and the magnetization direction, *θ*. Its value changed from 82.4 degree for the non-irradiated to 77.2 degree for the irradiated sample of $Fe_{94.2}Cr_{5.8}$, from 81.7 degree to 75.4 degree, respectively for the $Fe_{89.25}Cr_{10.75}$, and from 64.7 degree to 51.4 degree for the most concentrated sample. The number of Cr atoms within the *1NN-2NN* volume has remained hardly effected in the least-concentrated sample as well as that in the medium-concentrated sample, and it decreased from 2.217 to 2.049 in the most-concentrated sample.

The actual distribution of Fe/Cr atoms can be conveniently described in terms of short-range-order (SRO) parameters, $\alpha_i$. As outlined elsewhere [6], for the Mössbauer spectroscopic data, as in the present case, they can be defined locally i.e. for a given *(m,n)*, or as average for a particular neighbour shell viz. for *1NN*, *<$\alpha_1$>*, for *2NN*, *<$\alpha_2$>* or for *1NN-2NN*, *<$\alpha_{12}$>*. Here, the average SRO-parameters will be presented. They are defined as follows:

$$<\alpha_i> = \frac{<k>}{<k_r>} - 1 \qquad (1)$$

where $k = m, n, m+n$ for $i = 1, 2, 12$, respectively, and $<m>$ is the average number of Cr atoms in *1NN*, $<n>$ is that in *2NN*, and $<m+n>$ is that in *1NN−2NN* as determined from the analysis of the Mössbauer spectra. The three symbols with subscript *r* represent the same quantities but calculated for the random distribution.

Figure 3 shows values of $<\alpha_{12}>$ versus *x* for the three samples. For all non-irradiated samples $<\alpha_{12}>$ is negative which is indicative of clustering of Cr atoms. However, the amplitude of $<\alpha_{12}>$ decreases with *x* which means that for more concentrated samples the distribution of Fe/Cr atoms is more random. It is clear from the data presented in this plot that the irradiation had the most profound effect on the most-concentrated sample for which a change of the sign of $<\alpha_{12}>$ is observed i.e. in the irradiated $Fe_{84.85}Cr_{15.15}$ sample there is an ordering. In other

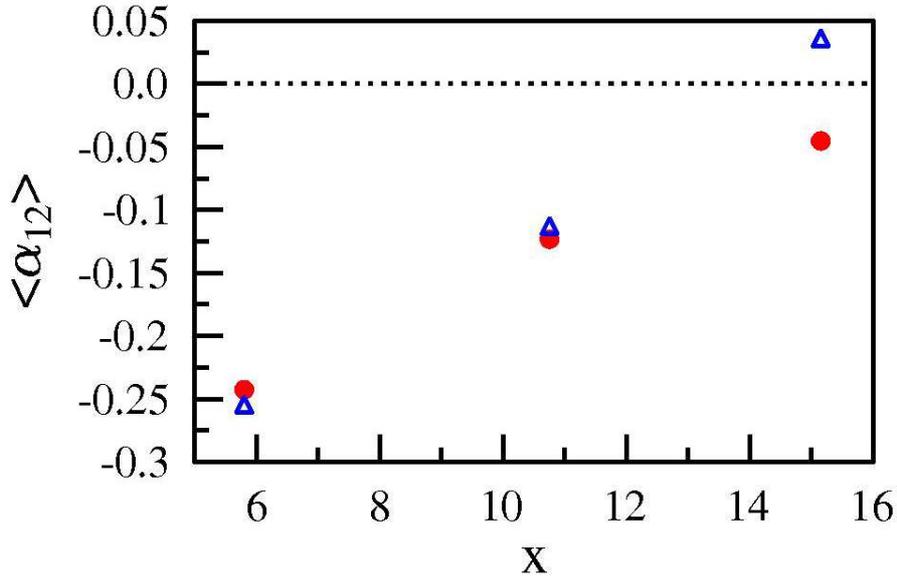

**Fig. 3** SRO $<\alpha_{12}>$ parameter versus *x* for non-irradiated (circles) and He-ions irradiated (triangles) samples.

words, one observes here an irradiation−induced inversion of $<\alpha_{12}>$.

In view of these results, the $Fe_{84.85}Cr_{15.15}$ sample was chosen for the further study of the effect of the irradiation dose on the spectral paramaters and SRO parameters.

### 3. 2. 2. The sample of $Fe_{84.85}Cr_{15.15}$

Irradiation of the $Fe_{84.85}Cr_{15.15}$ samples with different doses enabled first studying the effect of the dose on the spectral parameters. The most effected ones turned out to be (a) the average hyperfine field, $<B>$, and (b) the intensity of the second (fifth) line in the sextet or the Clebsch-Gordon coefficient, $C2$. The relevant plots are illustrated in Fig. 4.

Concerning the average hyperfine field, its value increases in a saturation-like way with the dose, *D*. The maximum increase of $<B> \approx 0.64$ T and it corresponds to a decrease of Cr content by ~2.3 at% according to the $<B>=f(x)$ relationship as given elsewhere [8]. Such effect can be understood in terms of a Cr atoms clustering. A further evidence in favour of this interpretation comes from the average number of Cr atoms situated in the *1NN-2NN* volume, *<m+n>*, which can be calculated as $<m+n> = \sum_{m,n} P(m,n)(m+n)$ Namely, *<m+n>* decreases from 2.121 for the non-irradiated sample, to 2.005 for the one irradiated to 25 dpa. The decrease is accompanied by a corresponding increase of *<B>* from 28.8 T to 29.5 T. The relationship between *<B>* and *<m+n>* - presented in Fig. 5 – can be well described in terms of a linear function whose slope is equal to 3.02 T/Cr atom which is significantly more than expected for the simple-dilution behaviour (2.36 T/Cr atom). A similar linear relationship between the two quantities was previously found for the Fe-Cr alloys with different Cr concentration [8,9]. Based on these observations one can conclude that the clustering of Cr atoms caused by the irradiation with He ions, is equivalent to a decrease of chromium concentration as seen by the Fe atoms.

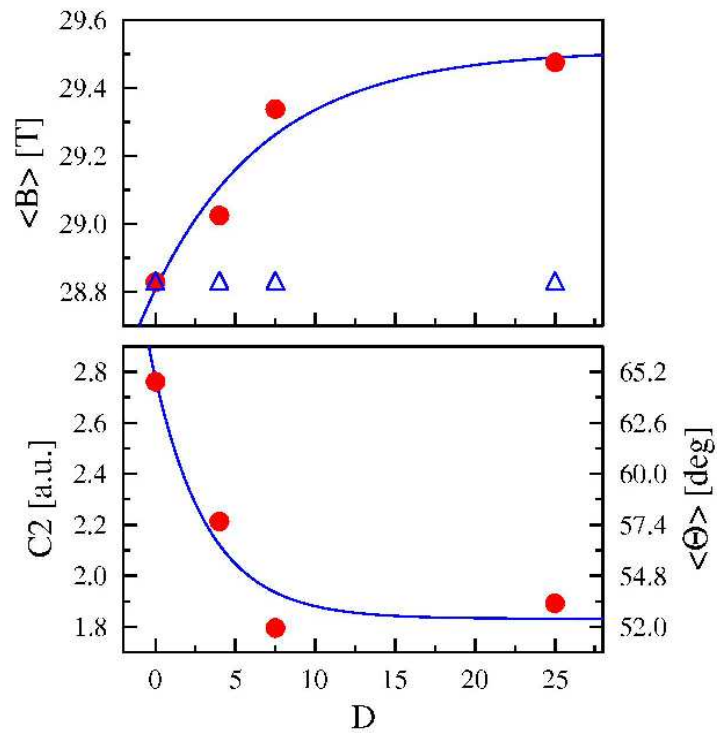

**Fig. 4** (Top) Dependence of the average hyperfine field, *<B>*, and (bottom) that of the relative intensity of the second (fifth) line in the sextet, *C2*, on the irradiation dose, *D*, as found for the $Fe_{84.85}Cr_{15.15}$ samples. The solid lines are guides to the eye to show the trends. In the top plot, *<B>* - values (triangles) expected for a random distribution of Fe/Cr atoms are added for comparison. The right-hand axis of the bottom-plot represents the corresponding average angle, *θ*, between the incident γ-rays and the magnetization vector. The sample represented by an open circle was irradiated in a second run.

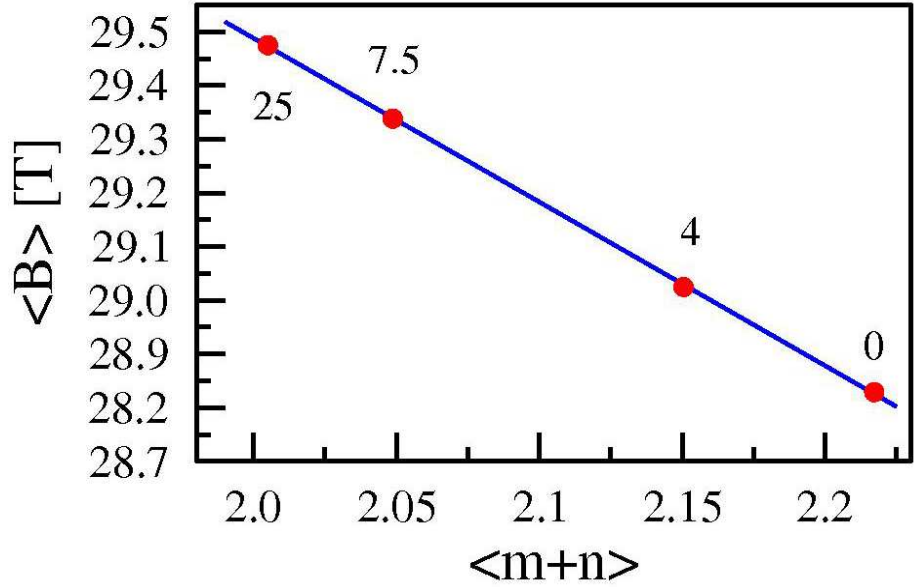

**Fig. 5** The average hyperfine field, $\langle B \rangle$, versus the average number of Cr atoms within the first two neighbour shells, $\langle m+n \rangle$, as determined for the $Fe_{84.85}Cr_{15.15}$ samples. The dose of irradiation is indicated for each point. The line represents the best linear fit to the data.

The second effect caused by the irradiation was a rotation of the magnetization vector towards the normal to the sample's surface. As shown in Fig. 4, an average angle, $\theta$, between the incident γ-rays that were perpendicular to the sample's surface and the magnetization vector has a value of $\langle \theta \rangle \approx 65$ degree for the non-irradiated sample, and it steeply decreases with $D$ in a saturation-like way. The maximum rotation's angle $\langle \theta \rangle \approx 13$ degree was achieved for $D \geq \sim 15$ dpa.

The observed effect of the irradiation on the hyperfine field could be explained in terms of the clustering of Cr atoms i.e. a re-distribution of Fe/Cr atoms must have taken place under the flux of He ions. The actual distribution of atoms within *1NN, 2NN and 1NN+2NN* shells can be, as outlined above, expressed in terms of the corresponding SRO−parameters $\langle \alpha_1 \rangle$, $\langle \alpha_2 \rangle$ and $\langle \alpha_{12} \rangle$, respectively.

A graphical presentation of $\langle \alpha_1 \rangle$, $\langle \alpha_2 \rangle$ and $\langle \alpha_{12} \rangle$ is displayed in Fig. 6 to illustrate the effect of the irradiation dose. It is clear that all three SRO parameters are different than zero which means the distribution of Fe/Cr atoms is not random. Concerning the *1NN*-shell, $\langle \alpha_1 \rangle$ is positive for all $D$-values which means that the number of Cr atoms situated in this shell is higher than the one expected for the random case. The overpopulation of Cr atoms in the *1NN*-shell means that Fe and Cr atoms being the first-nearest neighbours with each-other attract themselves. This situation is usually termed as ordering. Its degree increases with $D$ in a saturation-like way. A very different behaviour has been revealed for the *2NN*-shell. Here, $\langle \alpha_2 \rangle$ is negative and it hardly depends on $D$. The negative $\langle \alpha_2 \rangle$ indicates that the potential between Fe and Cr atoms being the second-nearest neighbours is repulsive leading to the underpopulation of Cr atoms in the *2NN*-shell. The overall effect i.e. the one observed within the *1NN-2NN* volume can be quantitatively expressed in terms of $\langle \alpha_{12} \rangle$. Its values, as shown in the top panel of Fig. 6, are much smaller than those of $\langle \alpha_1 \rangle$ and $\langle \alpha_2 \rangle$. This means that the departure from the random distribution is much smaller when measured as average over the two shells which is clear from the behaviour of $\langle \alpha_1 \rangle$ and $\langle \alpha_2 \rangle$. The SRO parameter averaged over the two shells shows a saturation-like increase from weakly negative values to weakly positive ones with an inversion at $D \approx 7$ dpa. The inversion is clearly caused by the behaviour observed in the *1NN*-shell, while the *2NN*-shell seems to be rather resistant to the applied irradiation.

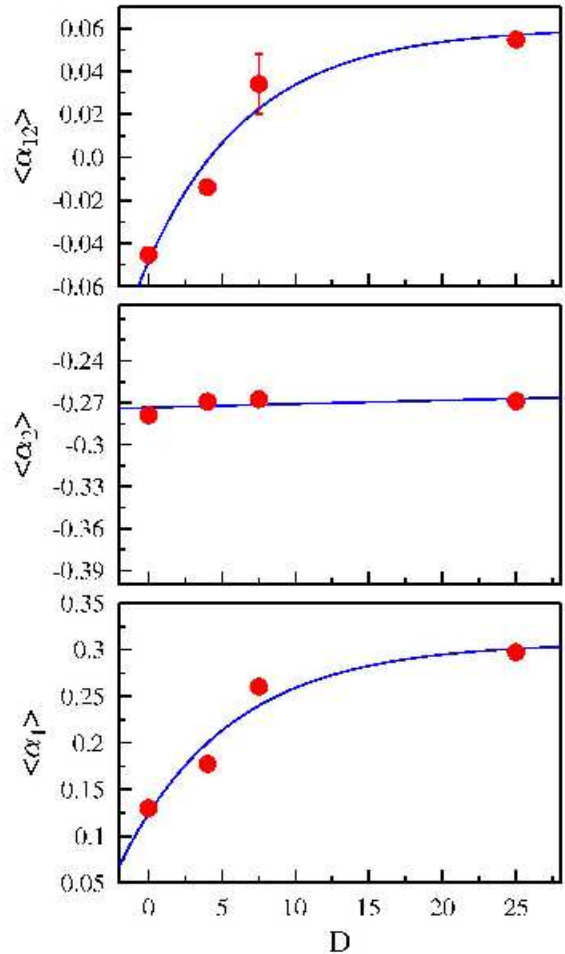

**Fig. 6** Dependence of the SRO-parameters $<\alpha_1>$, $<\alpha_2>$ and $<\alpha_{12}>$ on the irradiation dose, *D*, for the $Fe_{84.85}Cr_{15.15}$ samples. The solid lines are the guides to the eye to indicate the trends. The open circle is for the sample that was irradiated in a second run.

## 4. Conclusions

The results obtained in this investigation can be concluded as follows:

1. The irradiation of $Fe_{100-x}Cr_x$ alloys with $x$ = 5.8, 10.75 and 15.15 hardly affected them except that with $x$ =15.15 where an inversion of the short-range-order parameter averaged over the first two coordination shells, $<\alpha_{12}>$, from negative (clustering of Cr atoms) to positive (ordering of Cr atoms), was revealed.
2. The irradiation of the $Fe_{84.85}Cr_{15.15}$ samples with different dose, $4 \leq D \leq 25$ dpa, resulted in a change of the spectral parameters (average hyperfine field, $<B>$, and the intensity of the second/fifth line of the sextet, *C2*), as well as in the re-distribution of atoms.

3. The increase of $<B>$ is indicative of Cr atoms clustering; its maximum value is equivalent to ~2.3 at% decrease of Cr concentration within the *1NN-2NN* volume around the probe nuclei.
4. The maximum change of *C2* corresponds to a rotation of the magnetization vector by an angle of ~13 degree towards the normal to the sample's surface.
5. The average SRO parameter for the *1NN* shell, $<\alpha_1>$, is positive (hence indicative of the ordering of Cr atoms) and its amplitude increases with the irradiation dose, *D*, in a saturative-like way.
6. The average SRO parameter for the *2NN* shell, $<\alpha_2>$, is negative (hence indicative of the clustering of Cr atoms) and its amplitude hardly depends on *D*.
7. The average SRO parameter for the *1NN-2NN* shells, $<\alpha_{12}>$, changes its sign at $D \approx 7$ dpa from negative to positive, indicating thereby an irradiation-induced inversion.

**Acknowledgements**

This work was supported by the Association EURATOM-IPPLM and Ministry of Science and Higher Education, Warsaw, Poland.